\documentclass[prl,aps,twocolumn,amsmath,amssymb,showpacs]{revtex4}
\usepackage{graphicx}
\usepackage{epsfig}
\include{graphics}
\begin{document}
\title{Far-from-equilibrium  Ostwald ripening in electrostatically driven granular
powders}
\author{M.V.
Sapozhnikov$^{1,2}$, A. Peleg$^{3}$, B. Meerson$^{4}$, I.S.
Aranson$^{1}$, and K.L. Kohlstedt$^{5}$}

\affiliation{$^{1}$Argonne National Laboratory, 9700 S. Cass
Avenue, Argonne, IL 60439}

\affiliation{$^{2}$Institute for Physics of Microstructures,
Russian Academy of Sciences, GSP-105, Nizhny Novgorod 603600,
Russia}

\affiliation{$^{3}$Theoretical Division, Los Alamos National
Laboratory, Los Alamos, NM 87545}

\affiliation{$^{4}$Racah Institute of Physics, Hebrew University
of Jerusalem, Jerusalem 91904, Israel}

\affiliation{$^{5}$Department of Physics and Astronomy, University
of Kansas, Lawrence, KS 66045}

\begin{abstract}
We report the first experimental study of  cluster size
distributions in electrostatically driven granular submonolayers.
The cluster size distribution in this far-from-equilibrium process
exhibits dynamic scaling behavior characteristic of the (nearly
equilibrium) Ostwald ripening, controlled by the attachment and
detachment of the ``gas" particles. The scaled size distribution,
however, is different from the classical Wagner distribution
obtained in the limit of a vanishingly small area fraction of the
clusters. A much better agreement is found with the theory of
Conti \textit{et al.} [Phys. Rev. E \textbf{65}, 046117 (2002)]
which accounts for the cluster merger.

\end{abstract}
\pacs{45.70.Mg, 45.70.Qj} \maketitle

Dynamics of big assemblies of macroscopic particles are poorly
understood, especially when inter-particle interactions are
strongly dissipative \cite{Jaeger,Kadanoff}. Additional
complications arise when the grain size goes below $0.1$ mm, and
non-trivial contact interactions come into play. As small
particles acquire an electric charge, the dynamics are governed by
the interplay between long-range electrostatic and short-range
contact forces. On the other hand, an efficient electrostatic
excitation of granular media becomes possible here, and offers new
opportunities for investigation and control.

Recently, a host of fascinating pattern-formation phenomena and
far-from-equilibrium phase separation processes have been observed
in electrostatically driven granular powders, in air or vacuum
\cite{Aranson,Howell,AMSV,Sapozhnikov1} and in poorly conducting
liquids \cite{Sapozhnikov2}. One of them is cluster coarsening
which is remarkably similar to the Ostwald ripening
\cite{LS,Wagner}, despite the fact that the system is very far
from equilibrium. This work deals with experiment and theory of
this process. We work with submonolayers of metallic powders,
i.e., the layer of particles only partially covers the bottom of
the experimental cell, and it is approximately one-particle high.
Conducting particles (we used $40\, \mu m$ spherical copper
particles, the total number of particles was about $10^7$) are
placed between the plates of a large plane capacitor, levelled
horizontally. We used $27\times 27$ cm transparent capacitor
plates (glass coated by indium doped tin oxide); the plate spacing
was $1.5 \,mm$. The DC electric field $E$ in the capacitor was
varied in the range $0-7\, kV/cm$. The experiments were performed
in the atmosphere of dry nitrogen to reduce adhesion of particles
on the plate due to humidity of air.  Visualization was achieved
with a CCD camera mounted above the cell.

\begin{figure}[ptb]
\includegraphics[width=8.5cm,clip=]{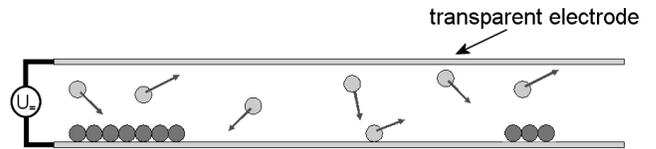}
\caption{A schematic of experimental setup.} \label{setup}
\end{figure}
When particles are in contact with the capacitor plate, they
acquire an electric charge. If the electric field in the cell $E$
exceeds a critical value, the resulting upward electric force
overcomes the particle weight and pushes the particles upward.
When a particle hits the upper plate, it recharges and falls back.
Remarkably, the collective behavior of this far-from-equilibrium
system closely resembles that of phase separating systems close to
equilibrium \cite{Aranson,Howell,AMSV,Sapozhnikov1}. The role of
temperature, however, is played by the electric field $E$. The
particles remain immobile on the bottom plate of the capacitor,
and form a precipitate phase, at $E<E_1$. If the field $E$ is
larger than a second threshold value, $E_2>E_1$, the particles
form a gas-like phase. This second field $E_2$ is 50\%-70\% larger
than $E_1$. Upon ``quenching" the field into the coexistence
region $E_1<E<E_2$ one observes nucleation of precipitate, and
small densely packed clusters form and start to grow on the bottom
plate. Then the clusters undergo coarsening: larger clusters grow
at the expense of smaller ones (which shrink and disappear). This
far-from-equilibrium coarsening process closely resembles the
Ostwald ripening with \textit{attachment-detachment} (in a
different terminology \textit{interface-controlled}) kinetics
\cite{Wagner}.

The basic physics behind phase separation and coarsening in this
system was investigated previously
\cite{Aranson,AMSV,Sapozhnikov1}. Two main physical processes are
at work here.  On the one hand, there is electrostatic screening:
a decrease in the vertical electric force $F$, exerted on a grain
in contact with the bottom plate, caused by the presence of other
grains.  On the other hand, collisions between the grains and the
bottom plate, and between the grains themselves, are strongly
inelastic, so the granular temperature is of no importance. The
reader is referred to Refs. \cite{Aranson,AMSV} for detail,
including quantitative estimates of the respective forces
\textit{etc}.

At the coarsening stage the clusters compete for the material, and
this competition is mediated by the gas phase. A preliminary
scaling analysis of experimental data on the cluster coarsening
dynamics \cite{Aranson} and special measurements of the shrinkage
rate of single clusters \cite{Sapozhnikov1} strongly indicated
that the cluster growth/coarsening is limited by the
attachment-detachment rates of the ``gas" particles, rather than
by the transport rate in the ``gas" phase. Correspondingly, a
phenomenological continuum model of this phase separation and
coarsening was formulated in Ref. \cite{AMSV} in terms of a
Ginzburg-Landau equation subject to conservation of the total
number of grains. Similar globally-conserved Ginzburg-Landau
equations have been studied in different contexts
\cite{Schimansky,MS,Conti}.  In the regime of well-developed
(densely-packed) clusters the continuum model yields
``sharp-interface" equations which govern the dynamics of the
inter-phase  boundary and are very useful in the analysis of the
late-time stage of the phase separation \cite{AMSV}.

This work focuses on the statistical dynamics of this
far-from-equilibrium Ostwald ripening process. One advantage of
our new experimental apparatus is a very large aspect ratio,
allowing us, for the first time, to produce a large number of
clusters and accumulate sufficient statistics.  By following the
dynamic scaling properties of the cluster size distribution we
establish with confidence the attachment-detachment character of
kinetics. Surprisingly, the scaled size distribution of the
clusters turns out to be quite different from the classical Wagner
distribution \cite{Wagner}. A much better agreement with
experiment is found when we employ a more advanced theory by Conti
\textit{et al.} \cite{Conti} which, in addition to the cluster
growth/shrinking processes, accounts for the cluster merger. The
theory has no adjustable parameters and is very different, both in
its assumptions and predictions, from a recent phenomenological
exchange-driven growth model suggested by Ben-Naim and Krapivsky
\cite{bennaim}.

\begin{figure}[ptb]
\begin{tabular}{cc}
\epsfxsize=4.0cm  \epsffile{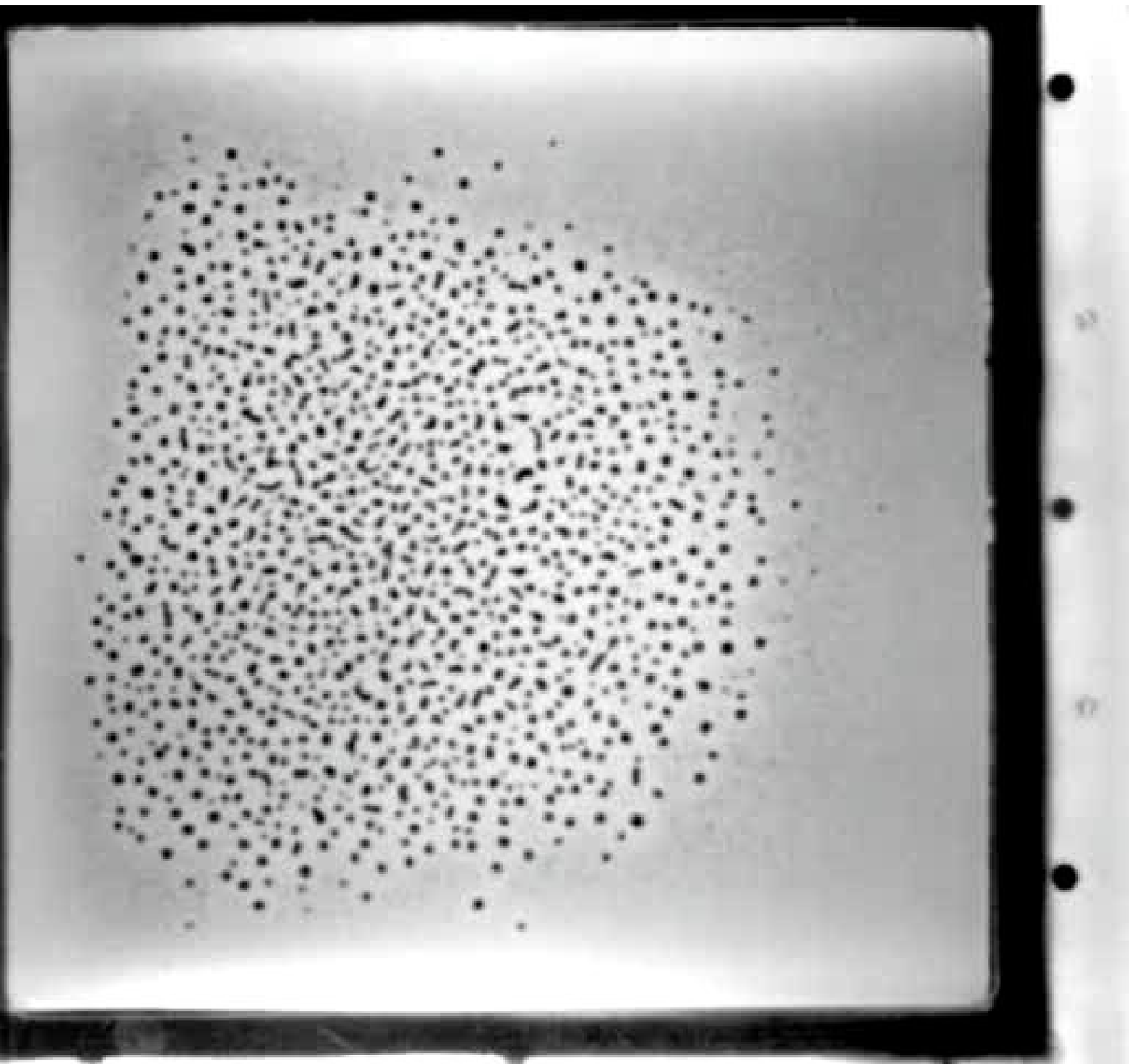} & \epsfxsize=4.0cm  \epsffile{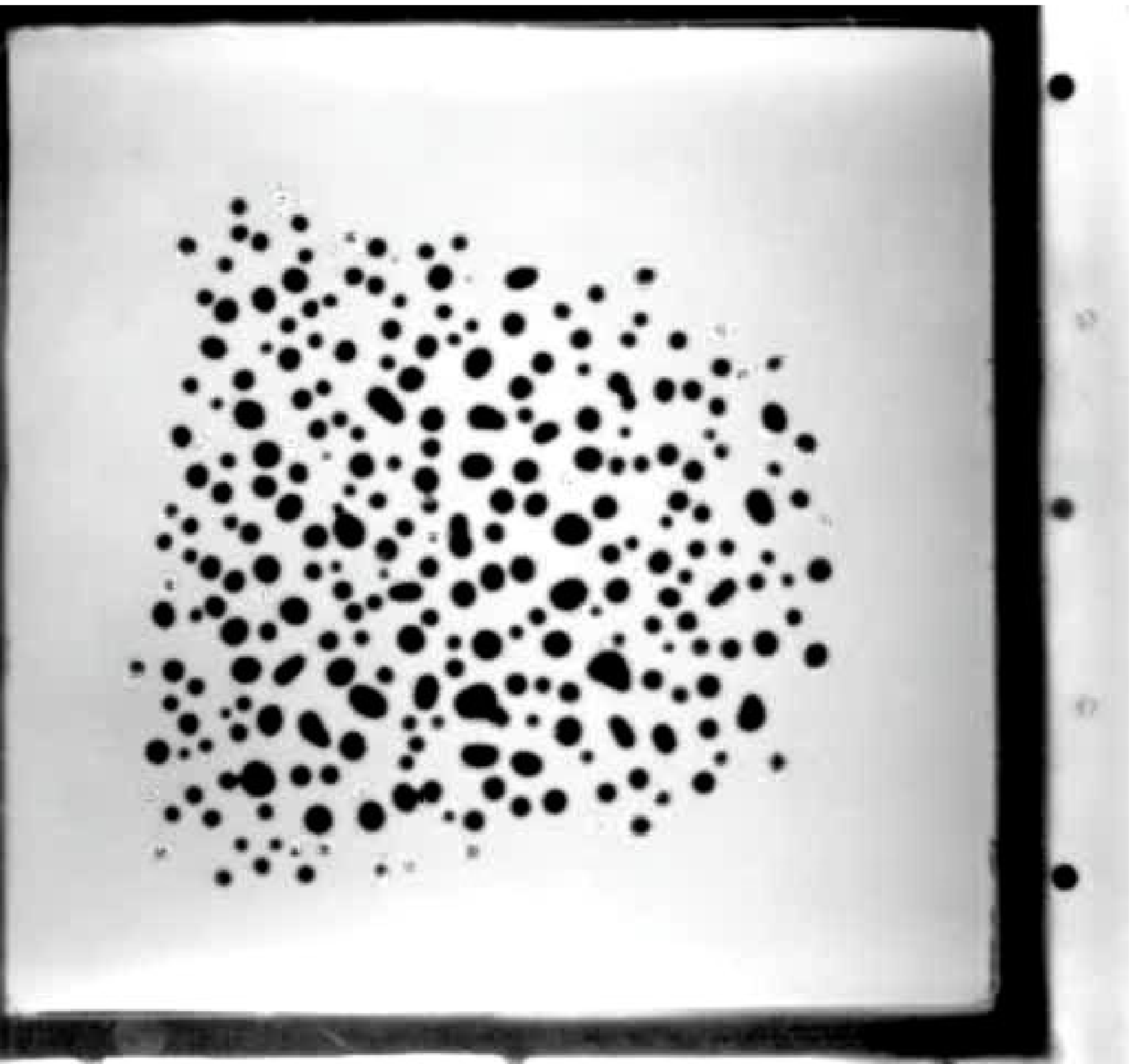}\\
\epsfxsize=4.0cm  \epsffile{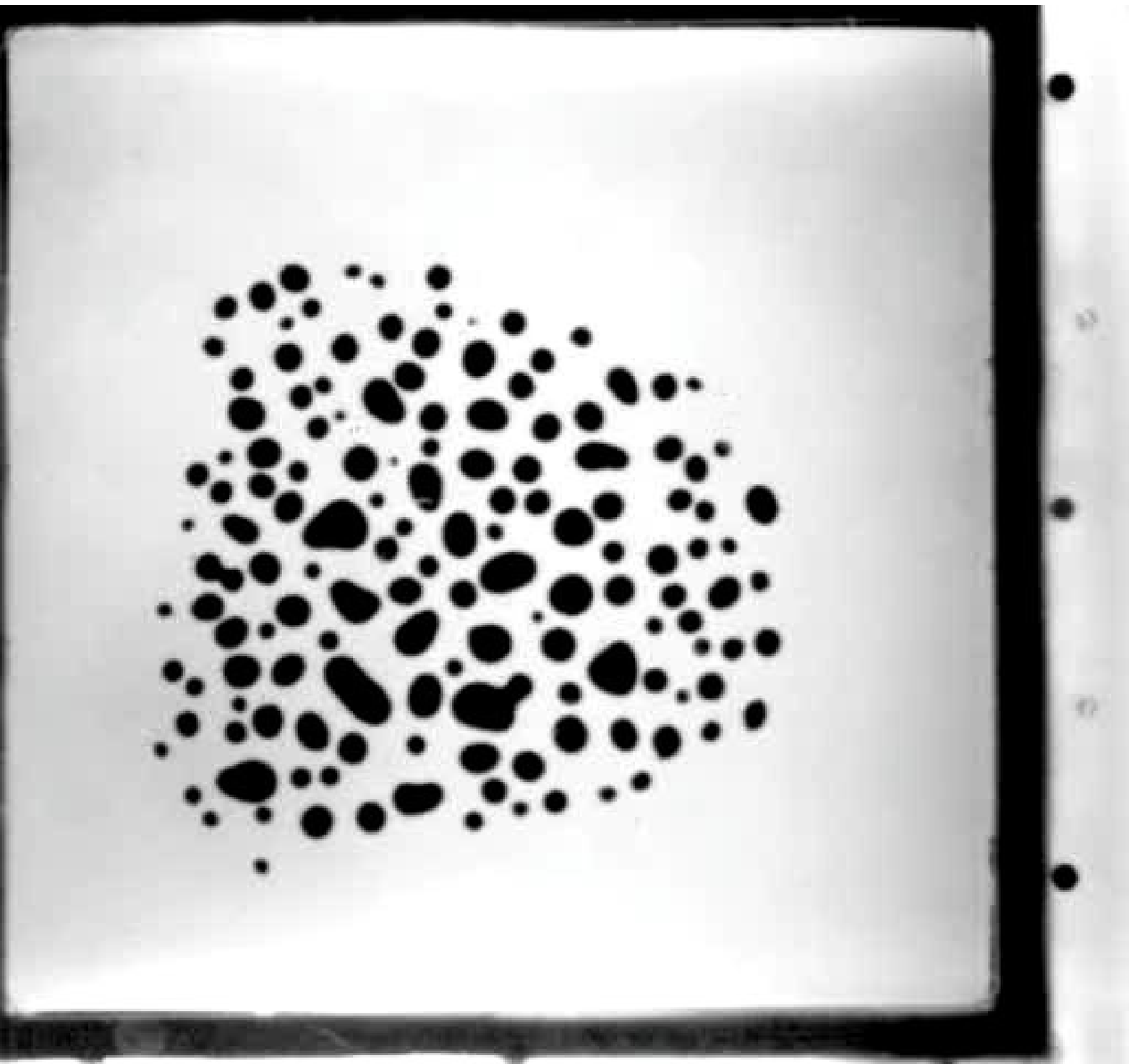} & \epsfxsize=4.0cm  \epsffile{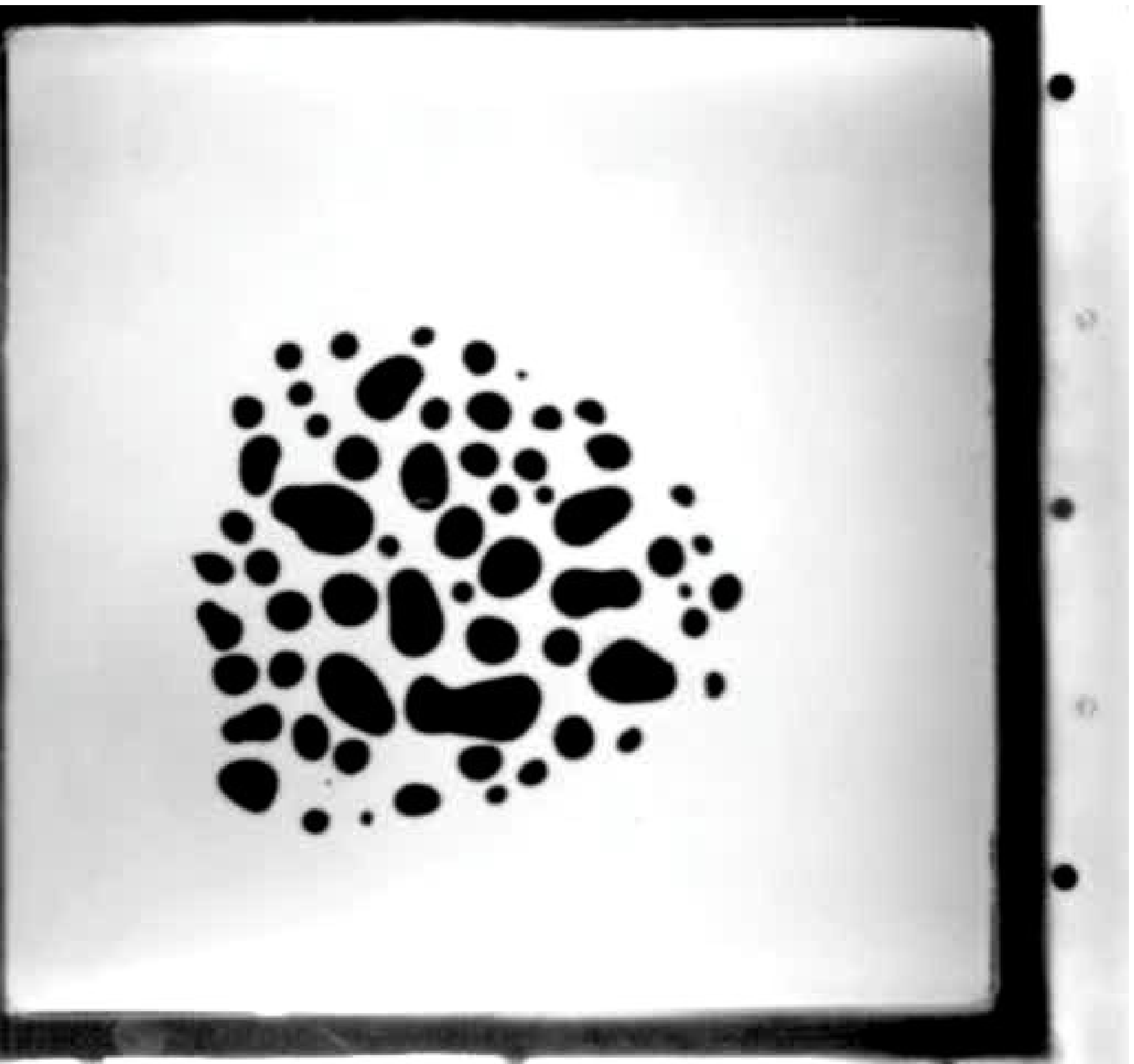}\\
\end{tabular}
\caption{Snapshots of coarsening of granular clusters in
electrostatic cell at times (in seconds) $t=0$ (upper left), $1
\times 10^4$ (upper right), $2 \times 10^4$ (lower left) and $5
\times 10^4$ sec (lower right). The applied DC electric field
$E=2.33$ kV/cm.} \label{fig2}
\end{figure}
Figure 2  shows a typical sequence of snapshots of the system in
the coexistence region $E_1<E<E_2$. The Ostwald ripening process
is clearly seen. In addition, distinct merger events are observed.
Figure 3a shows the average cluster area $\langle A \rangle (t)$
versus time. This dependence is very close to linear. Introducing
the cluster radius $R=(A/\pi)^{1/2}$, we find that the mean-square
radius $R_*(t)\equiv \langle R^2 \rangle^{1/2}$ grows with time
like $t^{1/2}$. This is a defining feature of the
attachment-detachment-controlled Ostwald ripening \cite{Wagner}.
The plot of the inverse number of clusters in the cell $N^{-1}$
versus time (Fig. 3b) approaches a straight line which implies a
scaling behavior $N(t) \sim t^{-1}$. An additional support for the
scaling is provided by the approximate constancy of the total area
of the clusters at late time, $N(t) \,R_*(t)^2 \sim const$, see
Fig. 4. The dynamic range of Figs. 3 and 4 is limited by
$5.40\times 10^{4}$, since for later times the number of clusters
is too small.

\begin{figure}[ptb]
\includegraphics[width=8.4 cm,clip=]{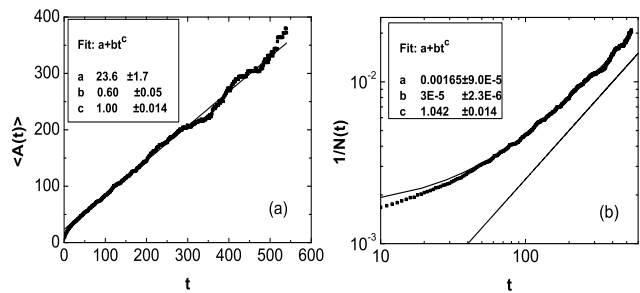}
\caption{The average cluster area $\langle A \rangle$ (a) and the
inverse number of clusters in the cell $1/N$ (b) versus time in
the experiment shown in Fig. 2. The unit of time in this figure
corresponds to $100 \, sec$. The solid curves (see also the
insets) show data fits of the type $a+b\,t^c$. The straight line
in figure b, shown for reference, has slope $1$. Notice that in
the experiments we waited for about two minutes (until small
clusters nucleated at the bottom plate) before starting the
measurements. As a result, the $\langle A \rangle (t)$ curve does
not pass through the origin.} \label{fig3}
\end{figure}
A more detailed characterization of the far-from-equilibrium
Ostwald ripening is provided by the dynamics of the probability
distribution function (PDF) of the cluster sizes $f(R,t)$. The PDF
is defined as usual:
\begin{equation}\label{f_definition}
\int_0^{\infty} f(R,t)\,dR=N(t)/L^2\,,
\end{equation}
where $L^2$ is the cell area. Dynamic scaling behavior of $f(R,t)$
is expected at late times, when the total area of the clusters
approaches a constant:
\begin{equation}\label{area}
\pi \int_0^{\infty} f (R,t) \,R^2  dR = \varepsilon\,.
\end{equation}
Here $\varepsilon$ is the area fraction of the clusters: the ratio
of the total area of all clusters to the cell area $L^2$. In view
of Eq. (\ref{area}), the expected scaling behavior is
\begin{equation}\label{scaled_DF}
f(R,t) = R_*^{-3}(t)\,F\left[R/R_*(t)\right]\,.
\end{equation}
Using the Ansatz (\ref{scaled_DF}), we can express the number of
clusters (\ref{f_definition}) and the mean squared radius
$R_*^2(t)$ through the moments of the scaled PDF $F(\xi)$:
\begin{eqnarray}
\label{clust_numb} \frac{N(t)}{L^2} = \frac{m_0}{R_*^2 (t)}\,,\\
\label{mean_square_1} R_*^2(t) \equiv \langle R^2 \rangle =
\frac{\int_0^{\infty} f(R,t) R^2\,dR}{\int_0^{\infty} f(R,t) \,dR}
= \frac{m_2}{m_0}\,R_*^2 (t)\,,
\end{eqnarray}
where $m_k=\int_0^{\infty}F(\xi)\,\xi^k d\xi\,, \,k=0,1,\dots$.
Equation (\ref{mean_square_1}) yields a consistency condition
$m_2=m_0$.

\begin{figure}[ptb]
\includegraphics[width=8.0cm,clip=]{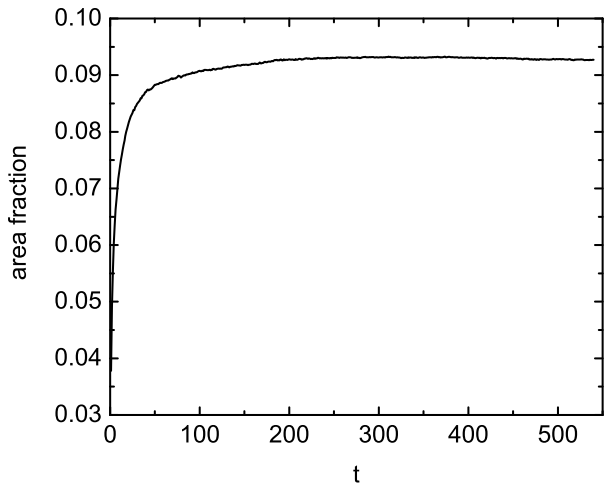}
\caption{The area fraction of the clusters (the total area of the
clusters divided by the area of the cell) versus time. The unit of
time corresponds to $100 \, sec$. The dynamic scaling regime is
observed at late times, when the area fraction approaches a
constant value.} \label{fig4}
\end{figure}
The imaging processing and tracking of the cluster growth and
coarsening in our experiment was performed using the MATLAB 6
toolbox.  The scaled PDF, obtained by averaging over 4 different
experiments, performed under the same conditions, is shown in Fig.
5. The horizontal axis shows the cluster radius divided by
$R_*(t)$, the vertical axis shows the experimental PDFs multiplied
by $R_*^3(t)$. The scaled PDF is the result of collapse of 340
original, unscaled PDFs from the 4 experiments, measured in the
time range $2.01 \times 10^4 \,\mbox{sec} \le t \le 5.40 \times
10^4$ sec with equal $100$ sec time intervals. We checked that the
late-time values of the cluster area fractions in these 4
experiments were close to each other: $9.1\%,\, 9.3\%$ (in two
experiments) and $9.4\%$.

\begin{figure}[ptb]
\includegraphics[width=8.0cm,clip=]{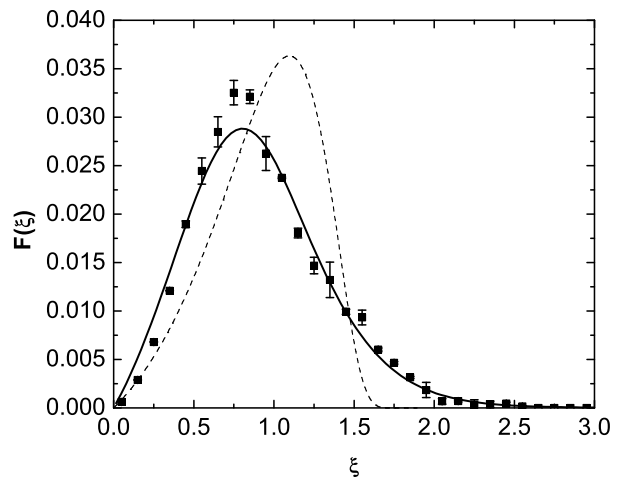}
\caption{Scaled probability distribution function (PDF) of cluster
sizes $F(\xi)$, where $\xi=R/R_*(t)$ [see Eq.
(\protect\ref{scaled_DF})]. The squares show the experimental
results, obtained from 340 snapshots from the interval $2.01
\times 10^4 \,\mbox{sec} \le t \le 5.40 \times 10^4$ sec and
averaged over 4 experiments. The error bars are not shown when
they are smaller than the size of the square. The dotted line is
the Wagner distribution $F_W (\xi)$ [see Eq.
(\protect\ref{wagner})], for the same area fraction
$\varepsilon=0.092$ as in the experiment. The solid line shows the
scaled PDF, obtained from the theory of the
attachment-detachment-controlled Ostwald ripening with coalescence
\cite{Conti} for the same area fraction.} \label{fig5}
\end{figure}
What is the theoretical prediction for the scaled PDF $F(\xi)$?
Let us first proceed by neglecting the cluster merger. In this
case the PDF dynamics is described by a continuity equation in the
space of cluster sizes:
\begin{equation}\label{C1}
  \partial_t f +\partial_R\, (\dot{R}f)=0\ .
\end{equation}
For the attachment-detachment-controlled kinetics, the cluster
growth/shrinking law is
\begin{equation}\label{Rdot}
  \dot{R} = D \left(\frac{1}{R_c(t)} - \frac{1}{R} \right)\,,
\end{equation}
where $R_c(t)$ is the time-dependent \textit{critical} radius,
while the effective diffusivity $D$ depends on the electric field
$E$ and on the size and weight of the grains \cite{AMSV}. The
growth law (\ref{Rdot}) is obtained under assumption that the
particle transport in the gas phase is very fast, so that the
concentration of the gas phase in the cell is approximately
uniform in space, varying only in time. The inverse critical
radius $1/R_c(t)$ in Eq. (\ref{Rdot}) is proportional to the
(time-dependent) difference between the concentration of the gas
phase and a constant value of this concentration, for which a
planar interface of the cluster phase is at rest \cite{AMSV}.
Equation (\ref{Rdot}) describes shrinking and disappearance of
small clusters, for which $R < R_c$. The shrinking occurs in a
finite time. At $R \ll R_c$ the area of the shrinking cluster goes
down linearly as the time increases. This property was checked in
experiment, and was found to hold until the shrinking clusters
become small compared to the gap between the capacitor plates, and
three-dimensional effects come into play \cite{Sapozhnikov1}.

Equations (\ref{C1}) and (\ref{Rdot}), together with the cluster
area conservation (\ref{area}), make a closed set. Mathematically,
this model is identical to the Wagner's model of the
attachment-detachment-controlled Ostwald ripening \cite{Wagner}.
(The attachment-detachment-controlled systems are quite different
from the \textit{diffusion}-controlled systems, for which a theory
of cluster coarsening, valid for vanishingly small area fractions
of the clusters, was developed earlier by Lifshitz and Slyozov
\cite{LS}.) An immediate consequence of Eqs. ({\ref{area}}),
(\ref{C1}) and (\ref{Rdot}) is the equality
$$
R_c(t) =\langle R \rangle (t) = \frac{\int_0^{\infty} f(R,t)
R\,dR}{\int_0^{\infty} f(R,t) \,dR}\,.
$$
For the scaling solutions (\ref{scaled_DF}) we obtain $R_c(t)
=\langle R \rangle (t) = (m_1/m_0) R_*(t)$. Therefore,  at large
times $\langle R \rangle (t) = const\, (Dt)^{1/2}$ and, in view of
Eq. (\ref{clust_numb}), $N(t)/L^2= const\, (Dt)^{-1}$, where the
prefactors depend on $\varepsilon$. Employing the Ansatz
(\ref{scaled_DF}) and solving the resulting ordinary differential
equation, one arrives at a \textit{family} of solutions for
$F(\xi)$, all of which have a compact support $0<\xi<\xi_m$. The
problem of selection of the correct scaled PDF out of this family
of solutions has been extensively studied
\cite{MS,AMS,GMS,Meerson}. A special role here is played by the
Wagner distribution,
\begin{equation}\label{wagner}
F_{W}(\xi)=\left\{\begin{array}{ll}
\frac{C \varepsilon \,\xi}{(\xi_m-\xi)^4}\, \exp\left(-\frac{2
\,\xi_m}{\xi_m -\xi}\right) & \mbox{if $0<\xi<\xi_m$}, \\
0 & \mbox{if $\xi>\xi_m$,}
\end{array}
\right.
\end{equation}
for which $\xi_m =\left[1+2 e^2\, Ei(-2) \right]^{-1/2} \simeq
1.8989$ is the largest. Also, $C=\pi^{-1}\left[1/(2 e^2) + Ei
(-2)\right]^{-1} \simeq 16.961\,,$ and $Ei \,(...)$ is the
exponential integral function \cite{Abramowitz}. The Wagner
distribution for $\varepsilon = 0.092$ (the same area fraction as
in experiment) is shown in Fig. 5. One can see a large discrepancy
between the Wagner distribution and the one observed in
experiment.

The Wagner's model is inaccurate because it disregards the cluster
merger. Therefore, we employed the more advanced theory of Conti
\textit{et al.} \cite{Conti} which accounts, within the framework
of the Ostwald ripening, for the binary coalescence of clusters
(growing clusters merge upon touching each other). Now Eq.
(\ref{C1}) gives way to the following kinetic equation:
\begin{eqnarray} \label{C6}
\partial_t f +\partial_R\, (\dot{R}f)  =
-\frac{1}{2}\int_0^\infty\,\int_0^\infty\,\{2M(R_1,R_2)\times
\nonumber\\
\left[\delta(R-R_1)+\delta(R-R_2)-\delta\left(R-\sqrt{R_1^2+R_2^2}\;\right)\right]\times
\nonumber\\
f(R_1,t)\, f(R_2,t)\, dR_1\, dR_2 \} \,,
\end{eqnarray}
where $M(R_1,R_2)=\pi\, (R_1+R_2)\, (\dot{R}_1+\dot{R}_2)\,
\theta(\dot{R}_1+\dot{R}_2)$, $\theta (\dots)$ is the step
function, and $\dot{R_1}$ and $\dot{R_2}$ are governed by the same
growth law (\ref{Rdot}). We refer the reader to Ref. \cite{Conti}
for a derivation of Eq. (\ref{C6}). Remarkably, Eqs. (\ref{area}),
(\ref{Rdot}) and (\ref{C6}) admit the same scaling Ansatz
(\ref{scaled_DF}). The scaled PDF is described by a nonlinear
integro-differential equation [which follows from Eq. (\ref{C6})],
subject to $m_2=m_0=\varepsilon/\pi$. The integro-differential
equation can be reduced to an integral equation and solved
iteratively \cite{Conti}. A crucial role in the iteration
procedure is played by the (non-normalizable) solutions of the
Wagner problem  which have an infinite support $0<\xi<\infty$. The
solution of the problem is unique (and normalizable), so the
account of merger resolves the selection problem intrinsic to the
Wagner's formulation. Figure 5 shows the scaled PDF, obtained by
the iteration procedure \cite{Conti} for the same area fraction
$\varepsilon = 0.092$ as in the experiment. One can see that the
agreement is much better than for the Wagner distribution.
Importantly, the only parameter which enters the theory of the
scaled PDF is the area fraction $\varepsilon$. Therefore, for a
given area fraction, a comparison of the theory and experiment
does not involve \textit{any} adjustable parameters. As can be
seen from Fig. 5, the theory works very well in describing the
position of the maximum of the scaled PDF and the shape of the PDF
to the right of the maximum, including the tail. On the other
hand, the theory overestimates the number of clusters at very
small scaled radii and underestimates it in the region of the
maximum of the PDF. This relatively small but systematic
disagreement may result from the assumption of an infinitely fast
transport in the gas phase. If this assumption is relaxed,
inter-cluster correlations will appear which should reduce the
effective merger rate. In our experiment, however, this reduction
is mitigated by the fact that, as time elapses, more material gets
concentrated in the middle of the cell (see Fig. 2). The future
work should attempt to account for these two competing effects.

In summary, we investigated the statistical dynamics of clusters
in electrostatically-driven granular powders. We found that a
recent advanced  theory of the attachment-detachment-controlled
Ostwald ripening \cite{Conti} yields a good quantitative
description of this \textit{far-from-equilibrium} system. It is
tempting to also apply this kind of analysis to mechanically
vibrated granular monolayers which exhibit strikingly similar
non-equilibrium phase separation properties
\cite{Sapozhnikov1,Urbach}.

We thank J.S. Olafsen and P.V. Sasorov for useful discussions.
This research was supported by the US DOE, Office of Basic Energy
Sciences (grant \# W-31-109-ENG-38) and by the Israel Science
Foundation (grant No. 180/02).

\end{document}